\documentclass[sigconf]{acmart}

\usepackage{booktabs} 
\usepackage{tabularx} 
\usepackage[english]{babel}
\usepackage{mathrsfs}

\usepackage{multirow}
\usepackage{multicol}
\usepackage{array}
\usepackage{arydshln}

\usepackage{amsmath}
\usepackage{amssymb}
\usepackage{booktabs}
\usepackage[inline]{enumitem}
\usepackage{siunitx}
\usepackage{stackengine}
\usepackage{extarrows}
\usepackage{graphicx}

\usepackage{caption}
\usepackage{subcaption}

\usepackage{footnote}
\makesavenoteenv{tabular}
\makesavenoteenv{table}

\usepackage{xspace} 

\newcommand{\specialcell}[2][c]{\begin{tabular}[#1]{@{}c@{}}#2\end{tabular}}

\newcolumntype{L}[1]{>{\raggedright\arraybackslash}p{#1}}
\newcolumntype{C}[1]{>{\centering\arraybackslash}p{#1}}
\newcolumntype{R}[1]{>{\raggedleft\arraybackslash}p{#1}}

\copyrightyear{2020}
\acmYear{2020}
\setcopyright{acmcopyright}
\acmConference[SIGIR '20]{Proceedings of the 43rd International ACM SIGIR Conference on Research and Development in Information Retrieval}{July 25--30, 2020}{Virtual Event, China}
\acmBooktitle{Proceedings of the 43rd International ACM SIGIR Conference on Research and Development in Information Retrieval (SIGIR '20), July 25--30, 2020, Virtual Event, China}
\acmPrice{15.00}
\acmDOI{10.1145/3397271.3401061}
\acmISBN{978-1-4503-8016-4/20/07}

\author{Helia Hashemi}
\affiliation{%
  \institution{University of Massachusetts Amherst}
}
\email{hhashemi@cs.umass.edu}

\author{Hamed Zamani}
\affiliation{%
  \institution{Microsoft}
}
\email{hazamani@microsoft.com}

\author{W. Bruce Croft}
\affiliation{%
  \institution{University of Massachusetts Amherst}
}
\email{croft@cs.umass.edu}

\newcommand{\model}{GT\xspace}

\begin{document}

\title{Neural Representation Learning for Clarification\\in Conversational Search}

\title{Guided Transformer:\\Leveraging Multiple External Sources for Representation Learning in Conversational Search}

\begin{abstract}
Asking clarifying questions in response to ambiguous or faceted queries has been recognized as a useful technique for various information retrieval systems, especially conversational search systems with limited bandwidth interfaces. Analyzing and generating clarifying questions have been studied recently but the accurate utilization of user responses to clarifying questions has been relatively less explored. In this paper, we enrich the representations learned by Transformer networks using a novel attention mechanism from external information sources that weights each term in the conversation. We evaluate this Guided Transformer model in a conversational search scenario that includes clarifying questions. In our experiments, we use two separate external sources, including the top retrieved documents and a set of different possible clarifying questions for the query. We implement the proposed representation learning model for two downstream tasks in conversational search; document retrieval and next clarifying question selection. Our experiments use a public dataset for search clarification and demonstrate significant improvements compared to competitive baselines.


\end{abstract}
\maketitle

\section{Introduction}
\label{sec:intro}
Conversational search has recently attracted much attention as an emerging information retrieval (IR) field. The ultimate goal of conversational search systems is to address user information needs through multi-turn natural language conversations. This goal is partially addressed in previous work with several simplifying assumptions. For example, the TREC Conversational Assistance Track (CAsT) in 2019 has focused on multi-turn conversational search, in which users submit multiple related search queries~\cite{Dalton:2020:CAST}. Similarly, conversational question answering based on a set of related questions about a given passage has been explored in the natural language processing (NLP) literature~\cite{Reddy:2019,Choi:2018,Qu:2019:CIKM}. However, the existing settings are still far from the ideal \emph{mixed-initiative} scenario, in which both user and system can take any permitted action at any time to perform a natural conversation. In other words, most existing work in conversational search assumes that users always ask a query and the system only responds with an answer or a ranked list of documents. 

Recent conversational information seeking platforms, such as Macaw~\cite{Zamani:2020:Macaw}, provide support for multi-turn, multi-modal, and mixed-initiative interactions. There have been recent efforts to go beyond the ``user asks, system responds'' paradigm by asking clarifying questions from the users, including offline evaluation of search clarification~\cite{Aliannejadi:CoRR:2019}, clarifying question generation for open-domain search queries~\cite{Zamani:2020:WWW}, and preference elicitation in conversational recommender systems~\cite{Christakopoulou:2016:KDD,Sepliarskaia:2018,zhang:CIKM:2018}. Past research in the area of search clarification has shown significant promise in asking clarifying questions. However, utilizing user responses to clarifying questions to improve the search performance has been relatively unstudied. In this paper, we propose a model that learns an accurate representation for a given user-system conversation. We focus on the conversations in which the user submits a query, and due to uncertainty about the query intent or the search quality, the system asks one or more clarifying questions to reveal the actual information need of the user. This is one of the many necessary steps that should be taken to achieve an ideal mixed-initiative conversational search system.

Motivated by previous research on improving query representation by employing other information sources, such as the top retrieved documents in pseudo-relevance feedback~\cite{Attar:1977,Croft:1988,Lavrenko:sigir:2001}, we propose a neural network architecture that uses multiple information sources for learning accurate representations of user-system conversations. We extend the Transformer architecture~\cite{Vaswani:CoRR:2017} by proposing a novel attention mechanism. In fact, the sequence transformation in Transformer networks are guided by multiple external information sources in order to learn more accurate representations. Therefore, we call our network architecture \emph{Guided Transformer} or \model. 
We train an end to end network based on the proposed architecture for two downstream target tasks: document retrieval and next clarifying question selection. In the first target task, the model takes a user-system conversation and scores documents based on their relevance to the user information need. On the other hand, the second task focuses on selecting the next clarifying question that would lead to higher search quality. For each target task, we also introduce an auxiliary task and train the model using a multi-task loss function. The auxiliary task is identifying the actual query intent description for a given user-system conversation. For text representation, our model takes advantage of BERT~\cite{Devlin:CoRR:Bert}, a state-of-the-art text representation model based on the Transformer architecture, modified by adding a ``task embedding'' vector to the BERT input to adjust the model for the multi-task setting. 

In our experiments, we use two sets of information sources, the top retrieval documents (similar to pseudo-relevance feedback) and the pool of different clarifying questions for the submitted search query. The rational is that these sources may contain some information that helps the system better represent the user information needs. We evaluate our models using the public Qulac dataset and follow the offline evaluation methodology recently proposed by \citet{Aliannejadi:CoRR:2019}. Our experiments demonstrate that the proposed model achieves over $29\%$ relative improvement in terms of MRR compared to competitive baselines, including state-of-the-art pseudo-relevance feedback models and BERT, for the document retrieval task. We similarly observe statistically significant improvements in the next clarifying question selection task compared to strong baselines, including learning to rank models that incorporate both hand-crafted and neural features, including BERT scores. 

In summary, the major contributions of this work include:
\begin{itemize}[leftmargin=*]
    \item Proposing a novel attention-based architecture, called Guided Transformer or \model, that learns attention weights from external information sources.
    \item Proposing a multi-task learning model based on \model for conversational search based on clarification. The multi-task learning model uses query intent description identification as an auxiliary task for training.
    \item Evaluating the proposed model on two downstream tasks in clarification-based conversations, namely document retrieval and next clarifying question selection.
    \item Outperforming state-of-the-art baseline models on both tasks with substantial improvements.
\end{itemize}

\section{Related Work}
\label{sec:rel}
In this section, we review related work on conversational search, search clarification, and neural model enhancement using external resources.

\subsection{Conversational Search and QA}
Although conversational search has become an emerging topic in the IR community in recent years, it has roots in early work on interactive information retrieval, such as \cite{Cool:1970:Expert,Croft:1987:inf,N.oddy:1997:CoRR}. For instance, \citet{Cool:1970:Expert} extracted how users can have an effective interaction with a merit information seeking system. Later on, \citet{Croft:1987:inf} introduced $I^3R$, the first IR model with a user modeling component for interactive IR tasks. Conversational system research in the form of natural language interaction started in the form of human-human interactions~\cite{Cool:1970:Expert} or human-system interactions with rule-based models~\cite{walker:2001:ACL}. Some early work also focus on spoken conversations in a specific domain, such as travel~\cite{Aust:1994:CoRR,Hemphill:1990:speech}.

More recently, \citet{radlinski:2017:CHIIR} introduced a theoretical framework and a set of potentially desirable features for a conversational information retrieval system. \citet{Tripas:2018:CHIIR} studied real user conversations and provided suggestions for building conversational systems based on human conversations. The recent improvements in neural models has made it possible to train conversation models for different applications, such as recommendation \cite{zhang:CIKM:2018}, user intent prediction \cite{Qu:2019:CHIIR}, next user query prediction~\cite{Yang:2017}, and response ranking \cite{Yang:2018:SIGIR}. 

There is also a line of research in the NLP community with a focus on conversational question answering \cite{Reddy:2019}. The task is to answer a question from a passage given a conversation history. In this paper, we focus on the conversations in which the system ask a clarifying question from the user, which is fundamentally different from the conversational QA literature.

\subsection{Asking Clarifying Questions}
Asking clarifying questions has attracted much attention in different domains and applications. To name a few, \citet{Pavel:2017:CHIIR} studied user intents, and clarification in community question answering (CQA) websites. \citet{Trienes:2019:ECIR} also focused on detecting ambiguous CQA posts, which need further follow-up and clarification. 
There is other line of research related to machine reading comprehension (MRC) task, that given a passage, generating questions which point out missing information in the passage. 
\citet{Rao:2019:NAACL} proposed a reinforcement learning solution for clarifying question generation in a closed-domain setting. We highlight that most of the techniques in this area assume that a passage is given, and the model should point out the missing information. Hence, it is completely different form clarifying the user information needs in IR. Clarification has also studied in dialog systems and chat-bots \cite{Boni:2003:NAACL, Boni:2005:NLE,lurcock:2004:Australian}, computer vision \cite{mostafazadeh:2016:ACL}, and speech recognition \cite{Stoyanchev:2014:CoRR}. However, since non of the above-mentioned systems are information seeking, their challenges are fundamentally different from challenges that the IR community faces in this area.

In the realm of IR, the user study done by \citet{Kiesel:2018:SIGIR} showed that clarifying questions do not cause user dissatisfaction, and in fact, they sometimes increase the satisfaction. \citet{Coden:2015:sumpre} studied the task of clarification for entity disambiguation. However, the clarification format in their work was restricted to a ``did you mean A or B?'' template, which makes it non-practical for many open-domain search queries. More recently, \citet{Aliannejadi:CoRR:2019} introduced an offline evaluation methodology and a benchmark for studying the task of clarification in information seeking conversational systems. They have also introduced a method for selecting the next clarifying question which is used in this paper as a baseline. \citet{Zamani:2020:WWW} proposed an approach based on weak supervision to generate clarifying questions for open-domain search queries. User interaction with clarifying questions has been later analyzed in~\cite{Zamani:2020:SIGIR}.

A common application of clarification is in conversational recommendation systems, where the system asks about different attributes of the items to reveal the user preferences. For instance, \citet{Christakopoulou:2016:KDD} designed an interactive system for venue recommendation. \citet{Sun:2018:CoRR} utilized facet-value pairs to represent a conversation history for conversational recommendation, and \citet{zhang:CIKM:2018} extracted facet-value pairs from product reviews automatically, and considered them as questions and answers. In this work, we focus on conversational search with open-domain queries which is different from preference elicitation in recommendation, however, the proposed solution can be potentially used for the preference elicitation tasks as well.

\subsection{Enhancing Neural Models using External Information Sources}
Improving the representations learned by neural models with the help of external resources has been explored in a wide range of tasks. Wu et al.~\cite{Wu:2018} proposed a text matching model based on recurrent and convolution networks that has a knowledge acquisition gating function that uses a knowledge base for accurate text matching. \citet{Yang:2018:SIGIR} studied the use of community question answering data as external knowledge base for response ranking in information seeking conversations. They proposed a model based on convolutional neural networks on top the interaction matrix. More recently, \citet{Yang:2019} exploited knowledge bases to improve LSTM based networks for machine reading comprehension tasks. Our work is different from prior work in multiple dimensions. From the neural network architecture, we extend Transformer by proposing a novel architecture for learning attention weights from external information sources. In addition, we do not use external knowledge bases in our experiments. For example, we use the top retrieved documents as one information source, which is similar to the pseudo-relevance feedback (PRF) methods~\cite{Attar:1977,Croft:1988,Lavrenko:sigir:2001,Zamani:2016:ICTIR}. We showed that our method significantly outperforms state-of-the-art PRF methods.



\section{Methodology}
\label{sec:method}
In this section, we first briefly review the basics of attention and self-attention in neural networks and then motivate the proposed solution. We further provide a detailed problem formulation, followed by the neural network architecture proposed in this work. We finish by describing the end to end multi-task training of the network.

\subsection{Background}
\label{sec:background}
Attention is a mechanism for weighting representations learned in a neural network. The higher the weight, the higher the attention that the associated representation receives. For example, \citet{Guo:2016} used IDF as a signal for term importance in a neural ranking model. This can be seen as a form of attention. Later on, \citet{Yang:2016} used a question attention network to weight the query terms based on their importance. Attention can come from an external source, or can be computed based on the representations learned by the network. 

The concept of attention has been around for a while. However, they really flourished when the self-attention mechanism has proven their effectiveness in a variety of NLP tasks~\cite{parikh:2016:CoRR,Paulus:2017:CoRR,Zhouhan:2017:CoRR,Vaswani:CoRR:2017,Devlin:CoRR:Bert}. Transformer networks~\cite{Vaswani:CoRR:2017} have successfully implemented self-attention and became one of the major breakthroughs in sequence modeling tasks in natural language processing. For instance, BERT \cite{Devlin:CoRR:Bert} is designed based on multiple layers of Transformer encoders and pre-trained for the language modeling task. Self-attention is a specific attention mechanism in which the representations are learned based on the attention weights computed by the sequence tokens themselves. In other words, the sequence tokens decide which part of the sequence is important to emphasize and the representations are learned based on these weights. The original Transformer architecture was proposed for sequence-to-sequence tasks, such as machine translation. The model consists of the typical encoder-decoder architecture. Unlike previous work that often used convolution or recurrent networks for sequence modeling, Transformer networks solely rely on the attention mechanism. Each encoder layer, which is the most relevant to this work, consists of a self-attention layer followed by two point-wise feed forward layers. The self-attention layer in Transformer is computed based on three matrices, the query weight matrix $W_{Q}$, the key weight matrix $W_{K}$, and the value weight matrix $W_{V}$. Multiplying the input token representations to these matrices gives us three matrices $Q$, $K$, and $V$, respectively. Finally the self-attention layer is computed as:
\begin{equation}
    Z = \text{softmax}(\frac{Q\times K^T}{\sqrt{d}})V
    \label{eq:self}
\end{equation}
where $d$ is the dimension of each vector. This equation basically represents a probability distribution over the sequence tokens using the $softmax$ operator applied to the query-key similarities. Then the representation of each token is computed based on the linear interpolation of values. To improve the performance of the self-attention layer, Transformer repeats this process multiple times with different key, query, and value weight matrices. This is called multi-headed self-attention mechanism. At the end, all $Z$s for different attention heads are concatenated as the output of multi-headed self-attention layer, and fed into the point-wise fully connected layer.

\subsection{Motivation}



In conversational search systems, users pursue their information needs through a natural language conversation with the system. Therefore, in case of uncertainty in query understanding or search quality, the system can ask the user a question to clarify the information need. A major challenge in asking clarifying questions is utilizing user responses to the questions for learning an accurate representation of the user information need. 

We believe that the user-system conversation is not always sufficient for understanding the user information need, or even if it is sufficient for a human to understand the user intent, it is often difficult for the system, especially when it comes to reasoning and causation. For example, assume a user submits the query ``migraines'', and the system asks the clarifying question ``Are you looking for migraine symptoms?'' and the user responds ``no!''. Although negation has been studied for decades in the Boolean information retrieval and negative feedback literature~\cite{Croft:2009,Peltonen:2017,Salton:1982,Wang:2008}, it is still difficult for a system to learn an effective representation for the user information need. 

In the above example, if external information sources cover different intents of query, the system can learn a representation similar to the intents other than ``symptoms''. We present a general approach that can utilize multiple different information sources for better conversation representation. In our experiments, we use the top retrieved documents (similar to the pseudo-relevance feedback assumption~\cite{Attar:1977,Croft:1988}) and all clarifying questions for the query as two information sources. Future work can employ user interaction data, such as click data, and past user interactions with the system as external sources.

\subsection{Problem Formulation}
\label{sec:problem formulation}



Let $Q = \{q_1, q_2, \cdots, q_n\}$ be the training query set, and $F_{q_{i}} =$ $\{f_{1q_{i}},$ $f_{2q_{i}},$ $\cdots,$ $f_{nq_{i}}\}$ denote the set of all facets associated with the query $q_i$.\footnote{The approaches are also applicable for ambiguous queries. Therefore, assume $F_{q_i}$ contains all aspects of the query.} In response to each query submitted by the user, a number of clarifying questions can be asked. Each conversation in this setting is in the form of $<q_i, c_1, a_1, c_2, a_2, \cdots, c_t, a_t>$, where $c_i$ and $a_i$ respectively denote the $i$\textsuperscript{th} clarifying question asked by the system and its answer responded by the user. The user response depends on the user's information need. The goal is to learn an accurate representation for any given conversation $<q_i, c_1, a_1, c_2, a_2, \cdots, c_t, a_t>$. The learned representations can be used for multiple downstream tasks. In this paper, we focus on (1) document retrieval and (2) next clarifying question selection.

\paragraph{Document Retrieval} In this task, each training instance consists of a user-system conversation with clarification, i.e., $<q_i,$ $c_1,$ $a_1,$ $\cdots,$ $c_t,$ $a_t>$, a document from a large collection, and a relevance label. 

\paragraph{Next Clarifying Question Selection} In this task, each training instance includes a user-system conversation with clarification (similar to above), a candidate clarifying question $c$ and a label associated with $c$. The label is computed based on the search quality after asking $c$ from the user.

Our model takes advantage of multiple information sources to better represent each user-system conversation. Let $S_{q_{i}} = \{s_{1{q_{i}}},$ $s_{2{q_{i}}},$ $\cdots,$ $s_{m{q_{i}}}\}$ denote a set of external information sources for the query $q_i$. Each $s_{j{q_{i}}}$ is a text. We later explain how we compute these information sources in our experiments. Note that the term ``external'' here does not necessary mean that the information source should come from an external resource, such as a knowledge graph. The term ``external'' refers to any useful information for better understanding of the user-system conversation that is not included in the conversation itself.

\begin{figure}[t]
    \centering
    \includegraphics[width=\linewidth,trim={0 3cm 14cm 8.5cm},clip]{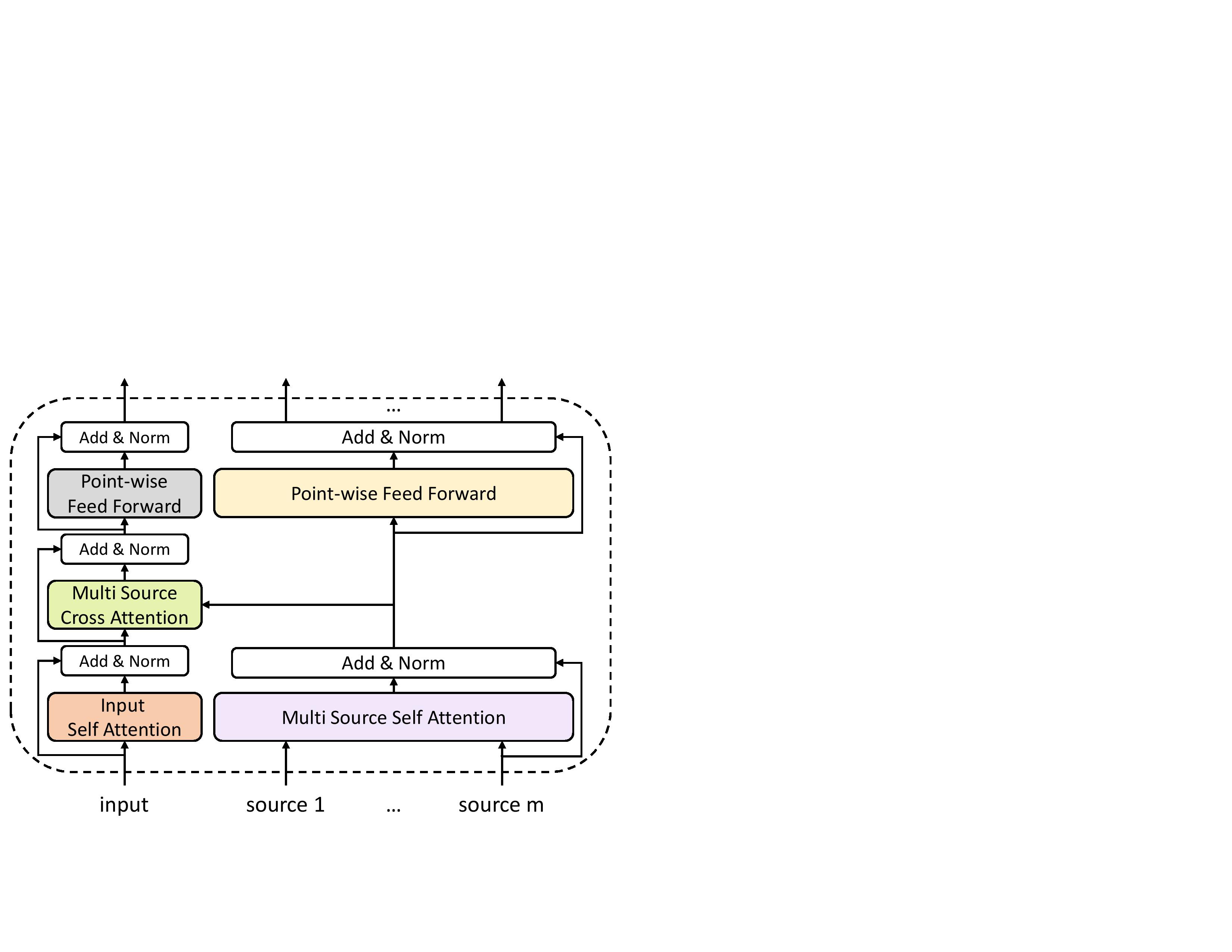}
    \caption{The architecture of each Guided Transformer layer on the input representations.}
    \label{fig:hhtransformer}
\end{figure}

\subsection{Guided Transformer}
\label{sec:net}



The architecture of each Guided Transformer (\model) layer is presented in \figurename~\ref{fig:hhtransformer}. The inputs of each \model layer is an input sequence (e.g., the user-system conversation) and $m$ homogeneous textual information sources. Each source is a set of sequences. We first feed the input sequence to a self-attention layer, called ``input self-attention''. This is similar to the self-attention layer in Transformer (see Section~\ref{sec:background}). In the self-attention layer, the representation of each token in the input sequence is computed based on the weighted linear interpolation of all token representations, in which the weights are computed based on the similarity of the query and key vectors, normalized using the softmax operator. See Equation~\ref{eq:self} for more details. We also apply a self-attention layer to the source representations. In other words, the ``multi source self-attention'' layer looks at all the sources and based on their similarity increases the attention on the tokens similar to those frequently mentioned in different source sequences. Based on the idea in residual networks~\cite{He:2016}, we add the input representations of each self-attention layer with its output and apply layer normalization~\cite{Ba:2016}. This is also the standard technique in the Transformer architecture~\cite{Vaswani:CoRR:2017}.

In the second stage, we apply attentions from multiple external sources to the input representations, i.e., the ``multi source cross attention'' layer. In this layer, we compute the impact of each token in external sources on each token in the input sequence. Let $t_i$ and $t_j^{(k)}$ respectively denote the $i$\textsuperscript{th} input token and the $j$\textsuperscript{th} token in the $k$\textsuperscript{th} source ($1 \leq k \leq m$). The output encoding of $t_i$ is computed as follows:
\begin{equation}
    \vec{t_i} = \sum_{k=1}^{m} \sum_{j=1}^{|s_k|} p_{\text{ca}}\left(t_j^{(k)} | t_i\right) \vec{v}_{t_j^{(k)}}
\end{equation}
where $s_k$ denotes the $k$\textsuperscript{th} external source and the vector $\vec{v}_{t_j^{(k)}}$ denotes the value vector learned for the token $t_j^{(k)}$. The probability $p_{\text{ca}}$ indicates the cross-attention weight. Computing this probability is not straightforward. We cannot simply apply a softmax operator on top of key-query similarities, because the tokens come from different sources with different lengths. Therefore, if a token in a source has a very high cross-attention probability, it would dominate the attention weights, which is not desired. To address this issue, we re-calculate the above equation using the law of total probability and the Bayes rule as follows:
\begin{equation}
\vec{t_i} = \sum_{k=1}^{m} \sum_{j=1}^{|s_k|} p\left(t_j^{(k)} | s_k, t_i\right) p\left(s_k | t_i\right) \vec{v}_{t_j^{(k)}}
\label{eq:ca-prob}
\end{equation}

In Equation~\ref{eq:ca-prob}, $p\left(t_j^{(k)} | s_k, t_i\right)$ denotes the attention weight of each token in source $s_k$ to the token $t_i$, and $p\left(s_k | t_i\right)$ denotes the attention weight of the whole source $s_k$ to the input token. This resolves the length issue, since $p\left(t_j^{(k)} | s_k, t_i\right)$ is normalized for all the tokens inside the source $k$, and thus no token in other sources can dominate the weight across all multi source tokens.

The cross-attention probability $p\left(t_j^{(k)} | s_k, t_i\right)$ is computed by multiplying the key vectors for the tokens in the source $s_k$ to the query vector of the token $t_i$, normalized using the softmax operator. To compute the attention probability $p\left(s_k | t_i\right)$, we take the key vector for the first token of the $s_k$ sequence. The rational is that the representation of the start token in a sequence represents the whole sequence~\cite{Devlin:CoRR:Bert}. Therefore, the multi source cross-attention layer can be summarized as follows in a matrix form:
\begin{equation}
    \sigma\left(\frac{Q' \times K^{'T}_{\text{\texttt{[CLS]}}}}{\sqrt{d}}\right) \sum_{k=1}^{m}{\sigma\left(\frac{Q \times K_k^T}{\sqrt{d}}\right)  V_k}
\end{equation}
where $K_k$ is the key matrix for the $k$\textsuperscript{th} external source, $Q$ is the query matrix for the input sequence, $K'_{\text{\texttt{[CLS]}}}$ is the key matrix for the first token of all sources, $Q'$ is another query matrix for the input sequence (using two separate of query weight matrices), $V_k$ is the value matrix for the $k$\textsuperscript{th} source, and $d$ is the dimension of the vectors. The function $\sigma$ is the softmax operator to transform real values in the $[-\infty, \infty]$ interval to a probabilistic space. Similar to the Transformer architecture, both self-attention and multi source cross-attention layers are designed based on the multi-headed attention, which is basically repeating the described process multiple times and concatenating the outputs. For the sake of space, we do not present the math for multi-headed attention and refer the reader to \cite{Vaswani:CoRR:2017}. Finally, the multi source cross-attention is followed by residual and layer normalization. Therefore, multiple \model layers can be stacked for learning more complex representations.

As shown in \figurename~\ref{fig:hhtransformer}, the last step in the multi source attention layer is a point-wise feed forward network. This is similar to the last step of the Transformer architecture, and consists of two point-wise feed forward layers with a ReLU activation in the first one. This feed forward network is applied to all tokens. A final residual and layer normalization produces the output of each multi source attention layer. The input and output dimension of multi source attention layers are the same. 

If there are different data with multiple instances as source signals, the model can be simply extended by adding more cross-attention layers, one per each data.

\begin{figure}[t]
    \centering
    \includegraphics[width=\linewidth,trim={0.25cm 0cm 9cm 8cm},clip]{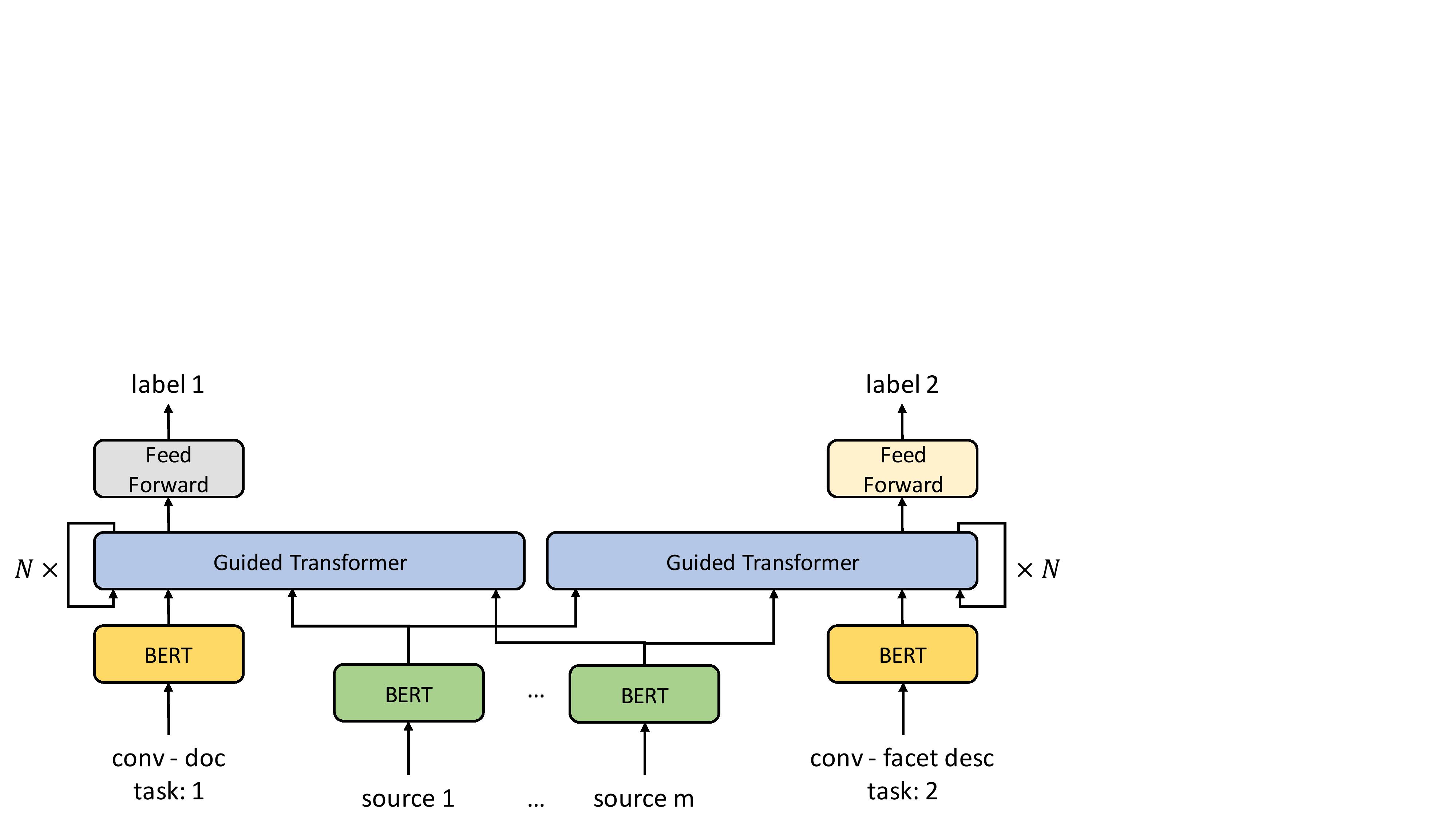}
    \caption{The high-level end to end architecture of the model trained using multi-task learning, where the first task is the target task (e.g., document ranking) and the second one is an auxiliary task that help the model identify the user information need from the user-system conversation. Same colors mean shared weights between the networks.}
    \label{fig:highlevel}
\end{figure}


\subsection{End to End Modeling and Training}
\label{sec:training}
The end to end architecture of the model is presented in \figurename~\ref{fig:highlevel}. As depicted in the figure, we use BERT for text representation. BERT~\cite{Devlin:CoRR:Bert} is a large-scale network based on Transformer which is pre-trained for a language modeling task. BERT has recently proven to be effective in a wide range of NLP and IR tasks, including question answering~\cite{Devlin:CoRR:Bert}, passage re-ranking~\cite{Nogueira:CoRR:2019,Padigela:2019}, query performance prediction~\cite{Hashemi:2019:ICTIR}, and conversational QA~\cite{Qu:2019:CIKM}. The coloring in \figurename~\ref{fig:highlevel} shows shared parameters. In other words, we use the same initial parameters for all the models, however, the parameters for BERTs with different colors are different and they are fine-tuned for accurate representation of their inputs. The output of external source representations and input representations are then fed to $N$ Guided Transformer layers introduced above. $N$ is a hyper-parameter. The representation for the start token of the input sequence (i.e., \texttt{[CLS]}) is then fed to a feed forward network with a single linear layer to predict the label. Note that for reusing BERT parameters in both tasks (the yellow components) we modified the BERT inputs, which is described later in this section.
    
As shown in the figure, we train our model using multi-task learning. The first task is the downstream target task (either document retrieval or next clarifying question selection) and the second task is an auxiliary task to help the model better learn the representation to identify the user information need. Note that we can train the model for three tasks (both target tasks and the auxiliary task), but due to GPU memory constraints we limit ourselves to two separate pairs of tasks. 

\paragraph{Task 1: The Target Task (Document Retrieval or Next Clarifying Question Selection)}
For the first task, we concatenate all the interactions in the conversation and separate them with a \texttt{[SEP]} token. For the document retrieval task, we concatenate the obtained sequence with the document tokens. For the next clarifying question selection task, we concatenate the obtained sequence with the next clarifying question. Therefore, the input of BERT for the document retrieval task is \texttt{[CLS] query tokens [SEP] clarifying question tokens [SEP] user response tokens [SEP] document tokens [SEP]}.\footnote{This example is only valid for a single clarifying question. The user-system conversation can contain more turns.} Note that BERT has a maximum sequence length limitation of $512$ tokens. Some documents in the collection are longer than this length limit. There exist a number of techniques to reduce the impact of this issue by feeding passages to the network and aggregating the passage scores. Since the focus of the work is on learning accurate conversation representations, we simply cut the documents that are longer than the sequence limit. 

Since we re-use the BERT parameters for the second task, we modify the BERT input by adding a \textbf{task embedding} vector. In other words, the model learns a representation for each task (in the multi-task learning setting) and we simply add the token embedding, positional embedding, the segment embedding and the task embedding. The first three embeddings are used in the original BERT model.

\paragraph{Task 2: The Auxiliary Task (Intent Description Identification)}
The clarifying questions are asked to identify the user information need behind the query. For example, each faceted query has multiple facets and each facet shows a query intent. Therefore, we used the intent description (or facet description) in the data (see Section~\ref{sec:problem formulation} for more details). Similar to the first task we concatenate the user-system conversation with the intent (or facet) descriptions. The label for this task is to identify whether the given facet description describes the user information need or not. In other words, for each user information need, we take some negative samples for training the network and the goal is to distinguish the correct query intent description. This auxiliary task helps the network adjust the parameters by learning attentions that focus on the tokens related to the relevant facet descriptions. Note that the intent description is only available at the training time and at the inference time we put the second part of the model (i.e., task 2) aside.

\paragraph{Loss Function} Our loss function is a linear combination of the target loss function and the auxiliary loss function:
\begin{equation}
    L = L_{\text{target}} + \alpha L_{\text{aux}}
    \label{eq:loss}
\end{equation}
where $\alpha$ is a hyper-parameter controlling the impact of the auxiliary loss function. Each of these loss functions is defined using cross entropy. For the task of document retrieval, the labels are binary (relevant vs. non-relevant), while, for the next clarifying question selection task the labels are real numbers in the $[0, 1]$ interval (which are computed based on the retrieval performance if the question is selected). Note that this loss function for the document ranking is equivalent to pointwise learning to rank. Previous work that uses BERT for text retrieval shows that point-wise BERT re-rankers are as effective as the pair-wise BERT models~\cite{Nogueira:CoRR:2019}.

\section{Experiments}
\label{sec:exp}
In this section, we evaluate the proposed model and compare against state-of-the-art baselines. First, we introduce the data we use in our experiments and discuss our experimental setup and evaluation metrics. We finally report and discuss the results.

\subsection{Data}
\label{sec:exp:data}

To evaluate our model, we use the recently proposed dataset for asking clarifying questions in information seeking conversations, called \textbf{Qulac} \cite{Aliannejadi:CoRR:2019}. Qulac was collected through crowdsourcing based on the topics in the TREC Web Track 2009-2012~\cite{Clarke:Trec:2009}. Therefore, Qulac contains 200 topics (two of which are omitted, because they have no relevant document in the judgments). Each topic has recognized as either ``ambiguous'' or ``faceted'' and has been also used for evaluating search result diversification. After obtaining the topics and their facets from the TREC Web Track data, a number of clarifying questions has been collected through crowdsourcing for each topic. In the next step, the authors ran another crowdsourcing experiment to collect answers to each clarifying question based on a topic-facet pair. The relevance information was borrowed from the TREC Web Track. The statistics of this dataset is reported in \tablename~\ref{tab:stats}. According to the table, the average facets per topic is $3.85 \pm 1.05$, and Qulac contains over 10k question-answer pairs.

The authors are not aware of any other IR dataset that contains clarifying questions.

\begin{table}[t]
    \centering
    \caption{Statistics of the Qulac dataset.}
    \label{tab:stats}
    \begin{tabular}{p{5cm}l}
        \toprule
         \# topics & 198 \\
         \# faceted topics & 141\\
         \# Ambiguous topics & 57\\
        \midrule

         \# facets & 762\\
         \# facet per topic & 3.85 $\pm$ 1.05 \\
         \# informational facets  & 577 \\
         \# navigational facets  & 185 \\
         \midrule

         \# clarifying questions  & 2,639 \\
         \#  question-answer pairs & 10,277\\
         \bottomrule
    \end{tabular}
    \vspace{-0.3cm}
\end{table}

\subsection{Experimental Setup}
\label{sec:exp:setup}

We use the language modeling retrieval model~\cite{ponte:1998:SIGIR} based on KL-divergence~\cite{Lafferty:2001} with Dirichlet prior smoothing~\cite{Zhai:2001} for the initial retrieval of documents from the ClueWeb collection. The smoothing parameter $\mu$ was set to the average document length in the collection. For document indexing and retrieval, we use the open-source Galago search engine.\footnote{\url{http://lemurproject.org/galago.php}} The spam documents were automatically identified and removed from the index using the Waterloo spam scorer\footnote{\url{https://plg.uwaterloo.ca/~gvcormac/clueweb09spam/}} \cite{Cormack:2011} with the threshold of $70\%$.

We evaluate the models using 5-fold cross-validation. We split the data based on the topics to make sure that each topic is either in the training, validation, or test set. To improve reproducibility, we split the data based on the remainder (the modulo operation) of the topic ID to $5$. Three folds are used for training, one fold for validation (hyper-parameter setting), and one fold for testing. After the hyper-parameters were selected based on the validation set, we evaluate the model with the seleted parameters on the test set. The same procedure was used for the proposed model and all the baselines.

We implemented our model using TensorFlow.\footnote{\url{https://www.tensorflow.org/}} We optimize the network parameters using the Adam optimizer~\cite{kingma:CoRR:2014} with the initial learning rate of $3 \times 10^6$, $\beta_{1} = 0.9$, $\beta_{2} = 0.999$, $L_{2}$ weight decay of $0.01$, learning rate warm-up over the first $5000$ steps, and linear decay of the learning rate. The dropout probability $0.1$ is used in all hidden layers. The number of attention heads in multi-headed attentions is set to $8$. The maximum sequence length for each document is set to $512$ (i.e., the BERT maximum sequence length). We use the pre-trained BERT-base model (i.e., 12 layer, 768 dimensions, 12 heads, 110M parameters) in all the experiments.\footnote{The pre-trained BERT models are available at \url{https://github.com/google-research/bert}} The batch size was set to $4$ due to memory constraints. The other hyper-parameters, including the parameter $\alpha$ (Equation~\ref{eq:loss}) and the parameter $N$ (the number of Guided Transformer layers), were selected based on the performance on the validation set.


\subsection{Evaluation Metrics}
\label{sec:exp:metrics}
Due to the nature of conversational search tasks, we focus on precision-oriented metrics to evaluate the models. We use mean reciprocal rank (MRR), and normalized discounted cumulative gain (nDCG)~\cite{Jarvelin:2002} with ranking cut-offs of @1, @5, and @20. We report the average performance across different conversations in the data. We identify statistical significant improvements using the paired t-test with Bonferroni correction at $95\%$ and $99\%$ confidence intervals (i.e., p-value less than $0.05$ and $0.01$, respectively).


\begin{table}[t]
    \caption{The retrieval performance obtained by the baseline and the proposed models. In this experiment, only one clarifying questions has been asked. $^\dagger$ and $^\ddagger$ indicate statistically significant improvements compared to all the baselines with $95\%$ and $99\%$ confidence intervals, respectively. $^*$ indicates statistical significant improvements obtained by MTL compared to the STL training of the same model at $99\%$ confidence interval.}
    \vspace{-0.3cm}
    \centering
    \setlength\tabcolsep{2.2pt}
    \begin{tabular}{lllllll}\toprule
         &\textbf{Method}  & & \textbf{MRR} & \textbf{nDCG@1} & \textbf{nDCG@5} & \textbf{nDCG@20} \\\midrule
        \multirow{5}{*}{\rotatebox[origin=c]{90}{\textbf{Baselines}}} &QL &&  0.3187 & 0.2127 & 0.2120 & 0.1866 \\
         &RM3  && 0.3196 & 0.2189 & 0.2149 & 0.2176 \\
         &ERM && 0.3291 & 0.2222 & 0.2191 & 0.2208 \\
         &SDM  && 0.3235 & 0.2267 & 0.2185 & 0.2182 \\
         &BERT  && 0.3527 & 0.2360 & 0.2267 & 0.2249  \\\midrule
         \multirow{7}{*}{\rotatebox[origin=c]{90}{\textbf{GT Network}}} & \textbf{Source} & \textbf{Loss} & \\
          & \multirow{2}{*}{Docs} & STL & 0.3710$^\ddagger$ & 0.2388 & 0.2309$^\dagger$ & 0.2284 \\
          & & MTL & 0.3826$^{\ddagger*}$ & 0.2407$^{\dagger*}$ & 0.2376$^{\ddagger*}$ & 0.2328$^{\dagger}$ \\\cdashline{2-7}
          & \multirow{2}{*}{CQs} & STL & 0.4028$^{\ddagger}$ & 0.2638$^{\ddagger}$ & 0.2521$^{\ddagger}$ & 0.2491$^{\ddagger}$ \\
          & & MTL & 0.4259$^{\ddagger*}$ & 0.2742$^{\ddagger*}$ & 0.2626$^{\ddagger*}$ & 0.2543$^{\ddagger*}$ \\\cdashline{2-7}
          & \multirow{2}{*}{\specialcell{Docs\\+CQs}} & STL & 0.4338$^{\ddagger}$ & 0.2792$^{\ddagger}$ & 0.2714$^{\ddagger}$ & 0.2610$^{\ddagger}$ \\
          & & MTL & \textbf{0.4554}$^{\ddagger*}$	& \textbf{0.2939}$^{\ddagger*}$ & \textbf{0.2803}$^{\ddagger*}$ & \textbf{0.2697}$^{\ddagger*}$  \\
         \bottomrule
    \end{tabular}
    \label{tab:results:main}
    \vspace{-0.5cm}
\end{table}

\subsection{Results and Discussion}
\label{sec:exp:results}

\subsubsection{Document Retrieval}
In the first set of experiments, we focus on conversations with only one clarifying question. The main reason behind this is related to the way the Qulac dataset was created. The questions in Qulac were generated by people operating in a realistic setting. However, the multi-turn setting is not as realistic as the single turn. Therefore, we first focus on single clarification in our main experiment and later extend it to multi-turn as suggested in \cite{Aliannejadi:CoRR:2019}. 

We compare \model with the following baselines:
\begin{itemize}[leftmargin=*]
    \item QL: The query likelihood retrieval model~\cite{ponte:1998:SIGIR} with Dirichlet prior smoothing~\cite{Zhai:2001}. The smoothing parameter was set to the average document length in the collection. This baseline also provides the first retrieval list for the re-ranking models.
    \item RM3: A state-of-the-art variant of relevance models for pseudo-relevance feedback~\cite{Lavrenko:sigir:2001}. We selected the number of feedback documents from $\{5, 10, 15, 20, 30, 50\}$, the feedback term count from $\{10, 20, \cdots, 100\}$, and the feedback coefficient from $[0,1]$ with the step size of $0.05$.
    \item ERM: Embedding-based relevance models that extends RM3 by considering word embedding similarities~\cite{Zamani:2016:ICTIR}. The ERM parameter range is similar to RM3.
    \item SDM: The sequential dependence model of \citet{Metzler:2005} that considers term dependencies in retrieval using Markov random fields. The weight of the unigram query component, the ordered window, and the unordered window were selected from $[0,1]$ with the step size of $0.05$, as the hyper-parameters of the model. We made sure that they sum to $1$.
    \item BERT: A pre-trained BERT model~\cite{Devlin:CoRR:Bert} with a final feed forward network for label prediction. This is similar to the method proposed by \citet{Nogueira:CoRR:2019}. BERT-base was used in this experiment, which is similar to the setting in the proposed model. The loss function is cross entropy. The optimizer and the parameter ranges are the same as to the proposed method. 
\end{itemize}

The hyper-parameters and training of all the models were done using 5-fold cross-validation, as described in Section~\ref{sec:exp:setup}. The same setting is used for the proposed methods. Note that the DMN-PRF model proposed by \citet{Yang:2018:SIGIR} was developed to use knowledge bases as external resources for response ranking in conversation. However, their model cannot accept long text, such as document level text, and thus we cannot use the model as a baseline. The machine reading comprehension models are all extracting answers from a passage and they require a passage as input, which is different from the setting in conversational search. Therefore, such models, e.g., \cite{Qu:2019:CIKM}, cannot be used as a baseline either. The BERT model has recently led to state-of-the-art performance in many IR tasks and is considered as a strong baseline for us.

We ran our model with different settings: single-task learning (STL) in which the only objective is the target task (i.e., document ranking) and multi-task learning (MTL) that optimizes two loss functions simultaneously. We also use the top 10 retrieved documents (i.e., Docs) and the top 10 clarifying questions (i.e., CQs) as external sources. 

\begin{table}[t]
    \centering
    \setlength\tabcolsep{3.2pt}
    \caption{Relative improvement achieved by \model with Docs+CQs and MTL compared to BERT for positive vs. negative user responses to clarifying question. $^*$ indicates statistical significant improvements at $99\%$ confidence interval.}
    \vspace{-0.3cm}
    \begin{tabular}{lcc}\toprule
        \textbf{Answer} & \textbf{\% MRR improvement} & \textbf{\% nDCG@5 improvement} \\\midrule
        Positive & 24.56\%$^*$ & 19.06\%$^*$ \\
        Negative & 31.20\%$^*$ & 25.84\%$^*$ \\\bottomrule
    \end{tabular}
    \label{tab:yes_no}
    \vspace{-0.3cm}
\end{table}

\begin{table}[t]
    \centering
    \setlength\tabcolsep{3.2pt}
    \caption{Relative improvement achieved by \model with Docs+CQs and MTL compared to BERT for user different response length to the clarifying question. $^*$ indicates statistical significant improvements at $99\%$ confidence interval.}
    \vspace{-0.3cm}
    \begin{tabular}{lcc}\toprule
        \textbf{Answer} & \textbf{\% MRR improvement} & \textbf{\% nDCG@5 improvement} \\\midrule
        Short & 33.12\%$^*$ & 26.65\%$^*$ \\
        Medium & 30.83\%$^*$ & 23.93\%$^*$ \\
        Long & 23.41\%$^*$ & 20.36\%$^*$ \\\bottomrule
    \end{tabular}
    \label{tab:response_length}
    \vspace{-0.5cm}
\end{table}

\begin{figure*}
\vspace{-0.4cm}
\begin{subfigure}{.25\textwidth}
  \centering
  \includegraphics[width=\linewidth]{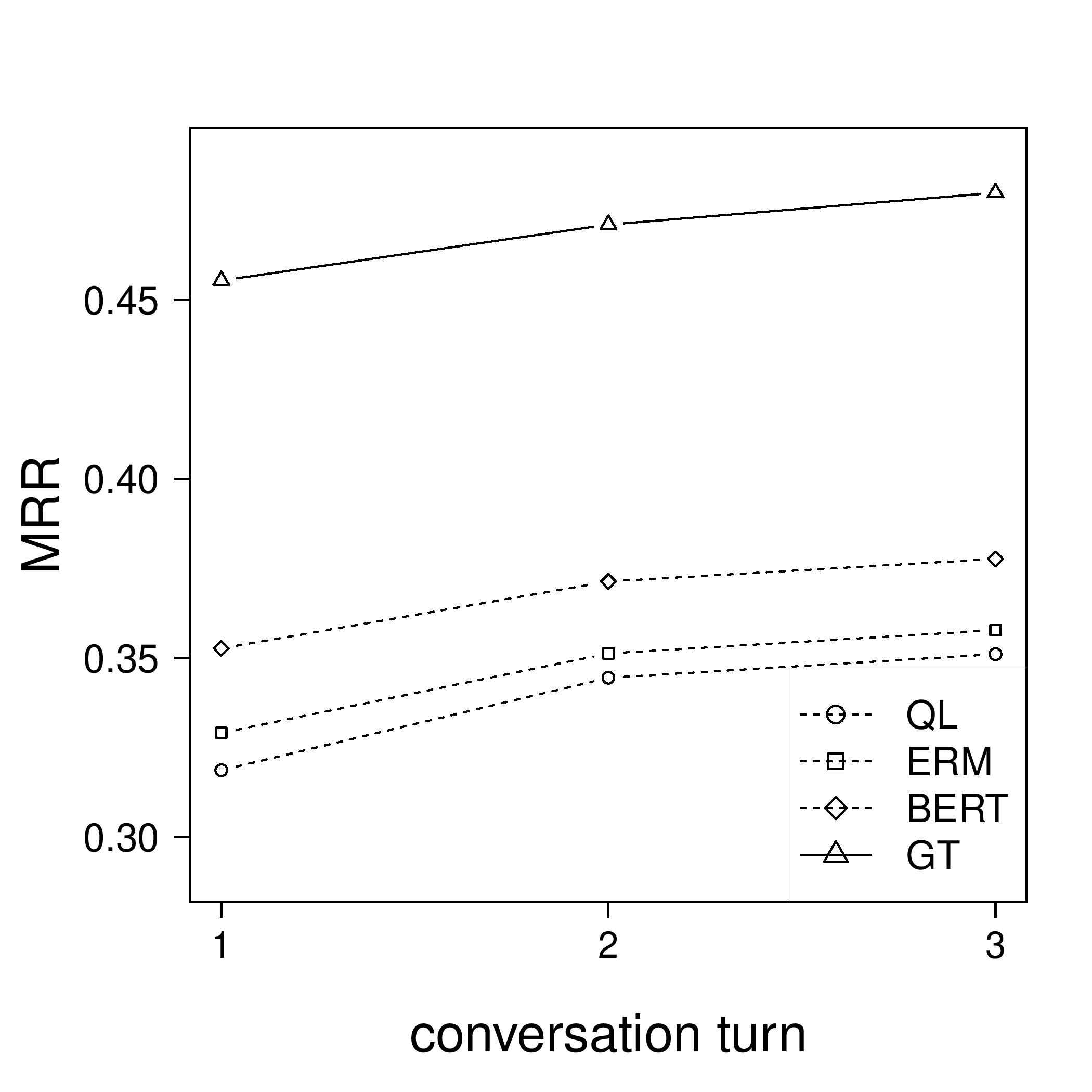}
\end{subfigure}%
\begin{subfigure}{.25\textwidth}
  \centering
  \includegraphics[width=\linewidth]{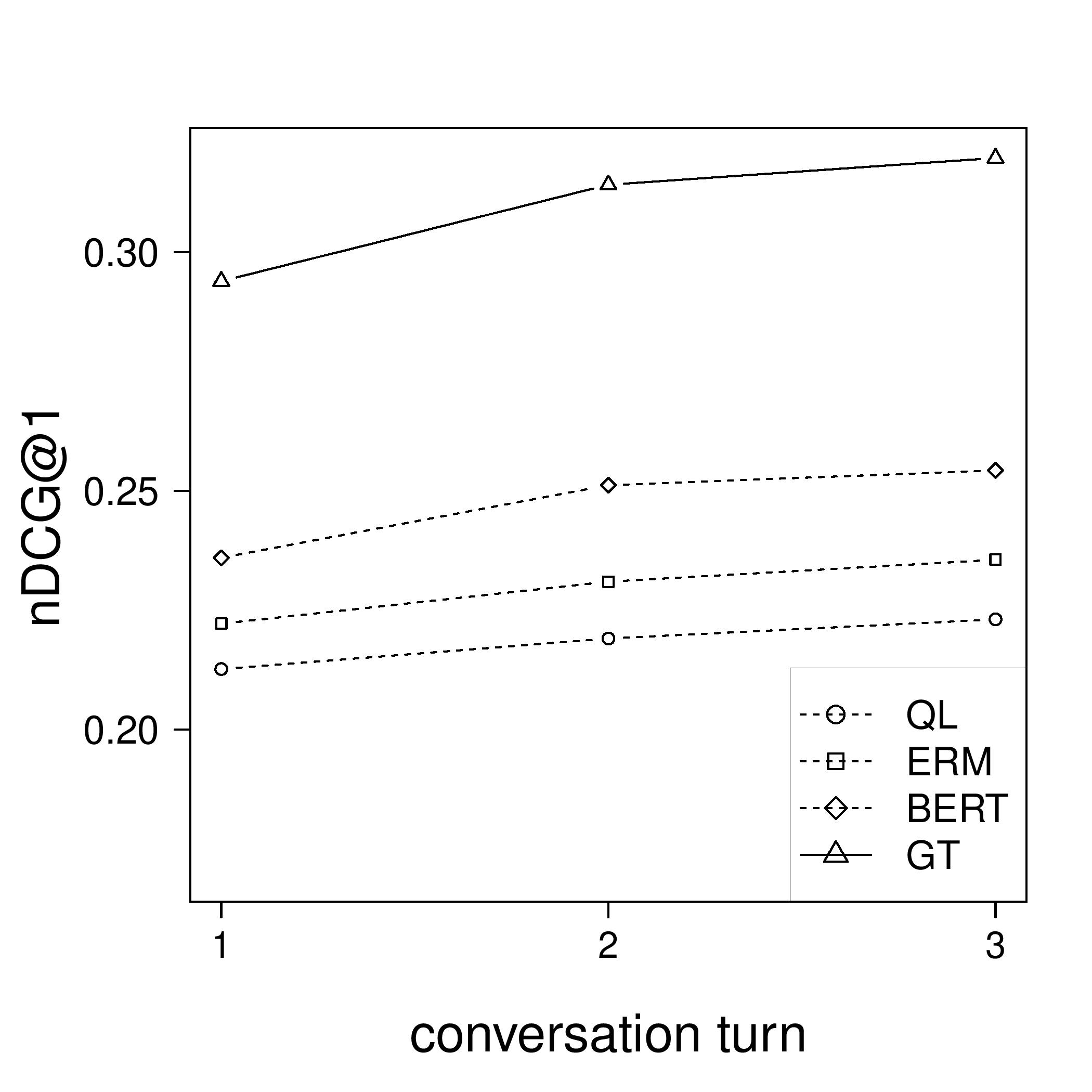}
\end{subfigure}%
\begin{subfigure}{.25\textwidth}
  \centering
  \includegraphics[width=\linewidth]{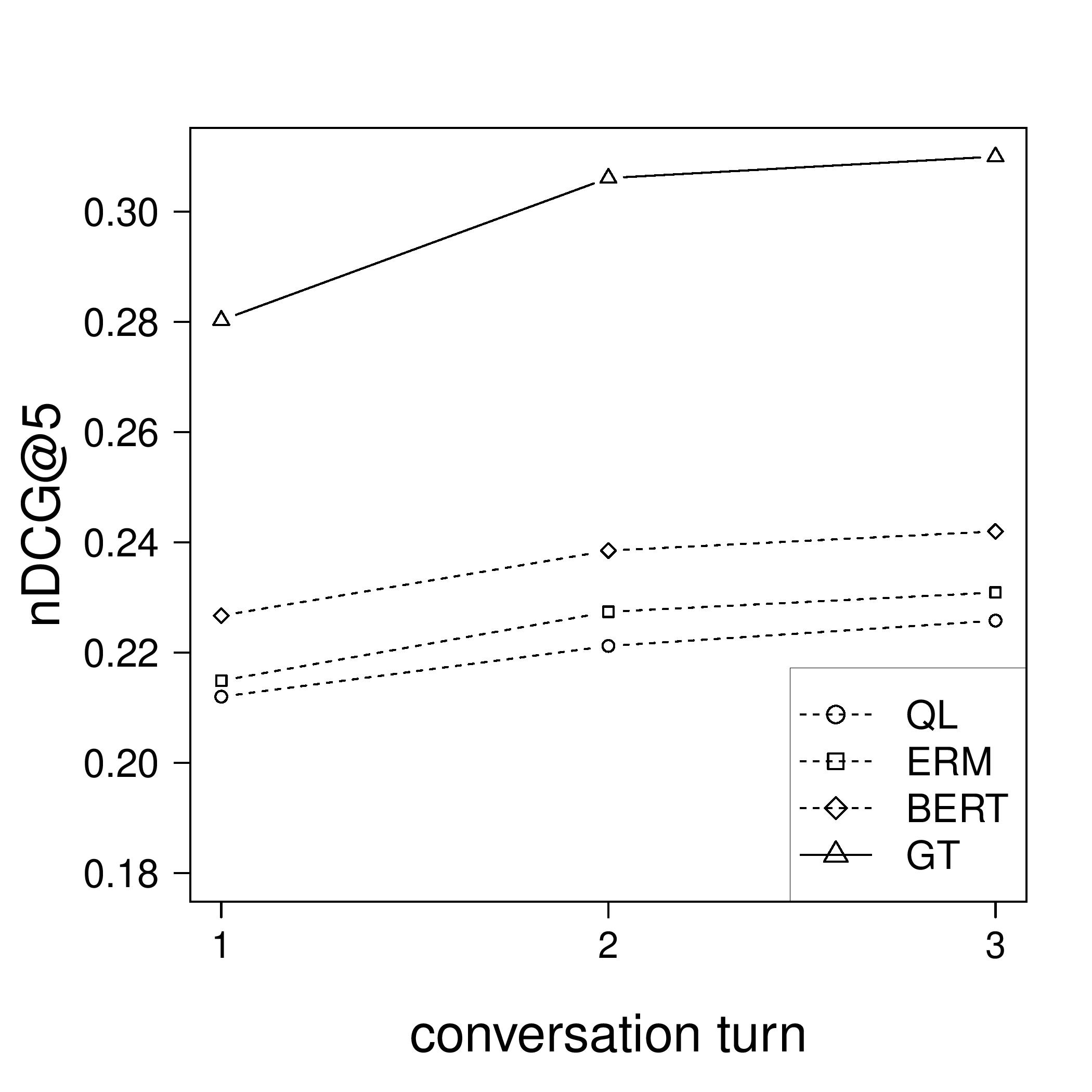}
\end{subfigure}%
\begin{subfigure}{.25\textwidth}
  \centering
  \includegraphics[width=\linewidth]{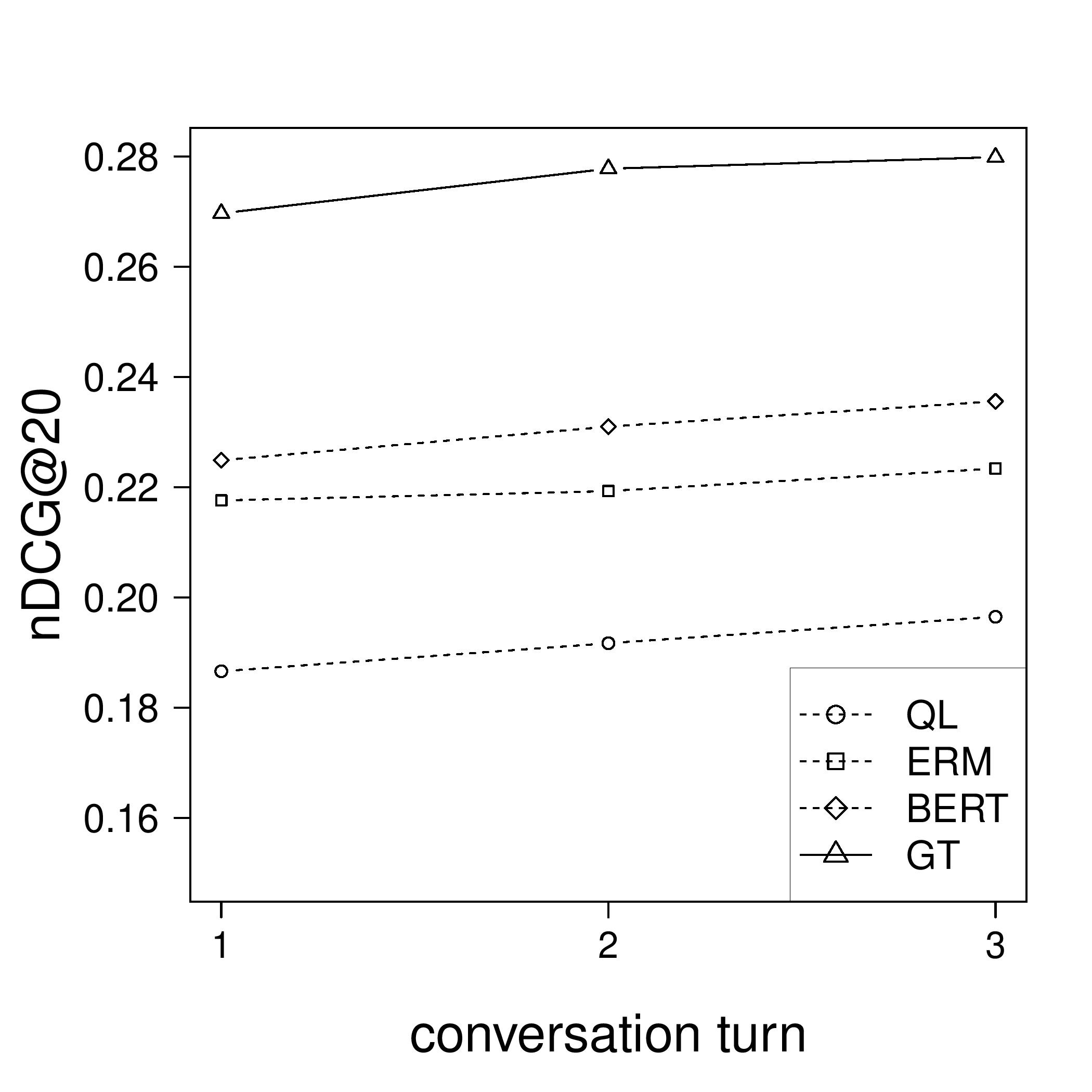}
\end{subfigure}%
\vspace{-0.3cm}
\caption{The performance of the \model (with Docs+CQs as source and MTL training) compared to the baselines for different conversation turns.}
\label{fig:turns}
\vspace{-0.5cm}
\end{figure*}

The results are reported in \tablename~\ref{tab:results:main}. According to the table, the proposed models significantly outperform all the baselines in nearly all cases. Using the clarifying questions as an external information source lead to a higher retrieval performance, compared to using the top retrieved documents. The reasons for this are two fold. First, the documents are often long and can be more than the 512 maximum sequence length. Therefore, the model cuts the documents and does not utilize all document tokens. Second, the top retrieved documents are not always relevant and non-relevant documents may introduce some noise. On the other hand, the clarifying questions are generated for the particular query and are all relevant. Note that although the clarifying questions in Qulac are human generated, recent methods are able to produce high quality clarifying questions for open-domain search queries. We refer the reader to \cite{Zamani:2020:WWW} for more details on automatic generation of clarifying questions.

According to the results, the multi-task learning setting outperforms the single-task learning setting, in all cases. This shows the effectiveness of using an auxiliary task for identifying the correct query intent description. Note that the facet descriptions are only used for training.

Taking both clarifying questions and the top retrieved documents as two information sources leads to the best performance in document retrieval. This shows that these two sources provide complementary information and the model can effectively utilize this information to better represent the user-system conversations.


\figurename~\ref{fig:turns} shows the performance curve as the conversation turn increases. Note that turn $i$ means that $i$ clarifying questions have been asked for each query. For the sake of visualization, we only plot QL, ERM, and BERT as the baselines and the \model with Docs+CQs as external source and MTL as the training setting. According to the plots, the performance of all models increases with the number of turns. However, the relative improvement in turn 3 compared to turn 2 is less than the improvements observed in turn 2 compared to turn 1. It is not practical to ask too many clarifying questions from users to answer their needs. The plots show it is also not helpful for the system. The proposed method substantially outperforms all the models in terms of all metrics, in all conversation turns.

To better understand the performance of the model, we report the results per different properties of user responses to clarifying questions. In this experiment, we only focus on a single clarifying question. We first identify yes/no questions (based on some simple regular expressions and the question response), and report the relative performance compared to BERT (our best baseline). For the sake of space, we only report the results for MRR and nDCG@5. The improvements for the other metrics also follow a similar behavior. For the proposed model, we focus on \model with the Docs+CQs source and MTL training. The results are presented in \tablename~\ref{tab:yes_no}. According to the table, \model achieved higher improvements for negative responses. The reason is that for positive responses, the clarifying question already contains some important terms about the user information need, and thus using external information sources leads to smaller improvements. This is true for all metrics, two of which are reported in the table.

We extend our analysis to the response length as well. In this experiment, we divide the test conversations into three equal size buckets based on the user response length to clarifying questions. The results are reported in \tablename~\ref{tab:response_length}. According to the table, \model achieves higher improvements for shorter user responses. This is due to the information that is in the responses. In other words, it is easier for BERT to learn an effective representation of user information need if enough content and context are provided, however, for shorter responses, external information sources are more helpful. In both Tables~\ref{tab:yes_no} and \ref{tab:response_length}, the improvements are statistically significant.

\begin{table}[t]
    \caption{Results for the next clarifying question selection task, up to 3 conversation turns. $^\dagger$ and $^\ddagger$ indicate statistically significant improvements compared to all the baselines with $95\%$ and $99\%$ confidence intervals, respectively. $^*$ indicates statistical significant improvements obtained by MTL compared to the STL training of the same model at $99\%$ confidence interval.}
    \vspace{-0.3cm}
    \setlength\tabcolsep{1.2pt}
    \centering
    \begin{tabular}{llllll}\toprule
         \textbf{Method}  & \textbf{MRR} & \textbf{nDCG@1} & \textbf{nDCG@5} & \textbf{nDCG@20} \\\midrule
         OriginalQuery & 0.2715 & 0.1381 & 0.1451 & 0.1470 \\
         $\sigma$-QPP  & 0.3570 & 0.1960 & 0.1938 & 0.1812 \\
         LambdaMART   & 0.3558 & 0.1945 & 0.1940 & 0.1796 \\
         RankNet    & 0.3573 & 0.1979 & 0.1943 & 0.1804 \\
         BERT-NeuQS     & 0.3625 & 0.2064 & 0.2013 & 0.1862 \\\midrule
         \model-Docs-STL & 0.3784$^\ddagger$ & 0.2279$^\ddagger$ & 0.2107$^\dagger$ & 0.1890   \\
         \model-Docs-MTL & \textbf{0.3928}$^{\ddagger*}$ & \textbf{0.2410}$^{\ddagger*}$ & \textbf{0.2257}$^{\ddagger*}$ & \textbf{0.1946}$^{\ddagger*}$   \\\midrule
         Oracle-Worst Question  & 0.2479 & 0.1075 & 0.1402 & 0.1483 \\
         Oracle-Best Question    & 0.4673 & 0.3031 & 0.2410 & 0.2077 \\
         \bottomrule
    \end{tabular}
    \label{tab:results:nextcq}
    \vspace{-0.5cm}
\end{table}

\begin{table*}[t]
    \centering
    \caption{Some examples with a single clarification turn. $\Delta$MRR is compared relative to the BERT performance.}
    \vspace{-0.3cm}
    \begin{tabular}{lll}\toprule
        Information need & Find sites for MGB car owners and enthusiasts. &  \\
        Query & mgb \\
        Clarifying question & are you looking for what mgb stands for? \\
        User response & no \\
        $\Delta$MRR & +78\% \\\midrule
        Information need & What restrictions are there for checked baggage during air travel? \\
        Query & air travel information \\
        Clarifying question & where are you looking to travel to? \\
        User response & doesn't matter i need information on checked baggage restrictions during air travel \\
        $\Delta$MRR & 3\% \\\midrule
        Information need & What states levy a tax against tangible personal property? &  \\
        Query & tangible personal property tax \\
        Clarifying question & would you like to find out how much you owe? \\
        User response & no i just want to know which states levy a tax against tangible personal property \\
        $\Delta$MRR & -21\% \\\bottomrule
    \end{tabular}
    \label{tab:examples}
    \vspace{-0.3cm}
\end{table*}

\subsubsection{Next Clarifying Question Selection}
In the second downstream task, we focus on next clarifying question selection. The task is to select a clarifying question given a user-system conversation. To be consistent with the results presented in \cite{Aliannejadi:CoRR:2019}, we use up to three conversation turns and report the average performance. The quality of models is computed based on the retrieval performance after asking the selected clarifying question from the user. Since the focus is on the next clarifying question selection, we use query likelihood as the follow up retrieval model, similar to the setting described in \cite{Aliannejadi:CoRR:2019}. 

We compare our method against the following baselines:
\begin{itemize}[leftmargin=*]
    \item OriginalQuery: The retrieval performance for the original query without clarifying question. We use this baseline as a term of comparison to see how much improvement we obtain by asking a clarifying question.
    \item $\sigma$-QPP: We use a simple yet effective query performance predictor, $\sigma$~\cite{sigmaqpp}, as an estimation of the question's quality. In other words, for a candidate clarifying question, we perform retrieval (without the answer) and estimate the performance using $\sigma$. The clarifying question that leads to the highest $\sigma$ is selected. 
    \item LambdaMART: We use LambdaMART to re-rank clarifying questions based on a set of features, ranging from the query performance prediction to question template to BERT similarities. The exact definition of feature descriptions can be found in \cite{Aliannejadi:CoRR:2019}. 
    \item RankNet: Another learning to rank model based on neural networks that uses the same features as LambdaMART. 
    \item BERT-NeuQS: A model based on BERT used for clarifying question re-ranking proposed in \cite{Aliannejadi:CoRR:2019}.
\end{itemize}

The hyper-parameters of these baselines were set using cross validation, similar to the proposed method. The results are reported in \tablename~\ref{tab:results:nextcq}. For the proposed method we only use the top retrieved documents as the external information source, since this is the most realistic setting for selecting clarifying questions. The proposed model significantly outperforms all baselines, including the recent BERT-NeuQS model. Consistent with the document retrieval experiments, the multi-task learning setting led to higher performance compared to STL. In addition, the achieved performance compared to the original query shows the benefit of asking clarification.

The table also contains the results for two oracle models, one that always selects the best question, which sets the upper-bound performance for this task on the Qulac data, and the one that always chooses the worst question (i.e., the lower-bound). The best possible performance is still much higher than the one achieved by the proposed solution and shows the potential for future improvement.

\subsubsection{Case Study}
We report three examples with one conversation turn for the document retrieval task in \tablename~\ref{tab:examples}. The $\Delta$MRR was computed relative to the BERT performance achieved by our best model. We report one win, tie, and loss example. In the first example, the user response is ``no'' and most baselines, including BERT, cannot learn a good representation from such conversation, however, the proposed model can use the top retrieved documents and the other clarifying questions to observe what are then other intents of the query and learn a better representation for the user information need. In the second example, the performance of the proposed model is similar to BERT, and in the last one, \model loses to BERT. The reason is that although the user response is negative, it contains some useful terms related to the information need, which makes it easier for the models to retrieve relevant documents. The proposed model, however, could not leverage external resources to learn an effective representation.


\section{Conclusions and Future Work}
\label{sec:conclusion}
In this paper, we introduced Guided Transformer (\model) by extending the Transformer architecture. \model can utilize external information sources for learning more accurate representations of the input sequence. We implemented \model for conversational search tasks with clarifying questions. We introduced an end to end model that uses a modified BERT representations as input to \model and optimizes a multi-task objective, in which the first task is the target downstream task and the second one is an auxiliary task of discriminating the query intent description from other query intents for the input user-system conversation. We evaluated the proposed model using the recently proposed Qulac dataset for two downstream tasks in conversational search with clarification: (1) document retrieval and (2) next clarifying question selection. The experimental results suggested that the models implemented using \model substantially outperform state-of-the-art baselines in both tasks. 

In the future, we intend to extend \model to a broad range of text understanding tasks using external sources. In the realm of conversational search, future work can focus on user past history as an external source for personalized conversational search. There are multiple different external sources that can use \model as an underlying framework for accurate representation learning for conversational search, such as query logs, clickthrough data, and knowledge bases. Most importantly, we are interested in extending this study to a fully mixed-initiative conversational search system, in which users can ask follow-up questions and at the same time the system can take different actions.

\paragraph{\textbf{Acknowledgement.}}
This work was supported in part by the Center for Intelligent Information Retrieval and in part by NSF IIS-1715095. Any opinions, findings and conclusions or recommendations expressed in this material are those of the authors and do not necessarily reflect those of the sponsor.



\end{document}